# Quasiparticle interaction originating from Bogoliubov Fermi Surfaces under pressure in 18%-S substituted FeSe studied via NMR

Zhongyu Yu[1], Xiaoling Shen[2,*], Koya Nakamura[1,†], Kazuya Inomata[1,§], Kohei Matsuura[3,‡], Yuta Mizukami[3,**], Shigeru Kasahara[4,††], Yuji Matsuda[4], Takasada Shibauchi[3], Yoshiya Uwatoko[2], and Naoki Fujiwara[1,§§]

**Abstract**
S-substituted FeSe superconductors in the tetragonal phase display several unique features among iron-based superconductors, particularly the presence of zero-energy excitations in the superconducting (SC) state. The recent concept of Bogoliubov Fermi Surfaces (BFSs)—a theoretical model describing ultranodal states—has attracted considerable interest. Nuclear magnetic resonance (NMR) studies on $FeSe_{1-x}S_x$ ($x=0.18$) have revealed an anomalous low-energy spin fluctuations deep in the SC state. The low-energy spin fluctuations are enhanced with decreasing temperature, supporting strong Bogoliubov quasiparticle interactions associated with BFSs. Here, we further investigate these correlation effects through $^{77}$Se-NMR measurements of $FeSe_{1-x}S_x$ ($x=0.18$) under pressures up to 2.0 GPa and temperatures down to ~100 mK. The results demonstrate that the anomalous enhancement is suppressed but persists under pressure, implying that quasiparticle interactions become weak by applying pressure. Furthermore, spin fluctuations in the normal state exhibit different temperature dependence from those deep in the SC state, suggesting that the nesting properties of normal electrons differ from those of Bogoliubov quasiparticles. These findings are consistent with the theoretical model of BFSs with $C_2$ symmetry and strengthen evidence for Bogoliubov quasiparticle interactions, providing insights into the unconventional pairing state of this system.

[1]Graduate School of Human and Environmental Studies, Kyoto University, Yoshida-Nihonmatsu-cho, Sakyo-ku, Kyoto, 606-8501, Japan
[2]Institute for Solid State Physics, University of Tokyo, 5-1-5 Kashiwanoha, Kashiwa, 277-8581, Chiba, Japan
[3]Department of Advanced Materials Science, University of Tokyo, 5-1-5 Kashiwanoha, Kashiwa, 277-8561, Chiba, Japan
[4]Department of Physics, Kyoto University, Kitashirakawa Oiwake-cho, Sakyo-ku, Kyoto, 606-8502, Japan
*Present address: Key Laboratory of Artificial Structures and Quantum Control, School of Physics and Astronomy, Shanghai Jiao Tong University, Shanghai 200240, China
†Present address: NEC Platforms, Ltd., 1753-1 Shimonumabe, Nakahara-ku, Kawasaki, Kanagawa, 211-8666, Japan
§Present address: DeNA Co., Ltd., 6-30-15 Hommachi, Shibuya-ku, Tokyo, 151-0071, Japan
‡Present address: Department of Applied Physics, University of Tokyo, 7-3-1 Hongo, Bunkyou-ku, Tokyo, 113-8656, Japan
**Present address: Graduate School of Science, Department of Physics, Tohoku University, Sendai, Miyagi, 980-8578, Japan
††Present address: Research Institute for Interdisciplinary Science, Okayama University, Okayama, 700-8530, Japan
§§Contact author: fujiwara.naoki.7e@kyoto-u.ac.jp

# INTRODUCTION

Iron selenide, FeSe is a unique superconductor among various iron-based superconductors in that it has extremely shallow unconnected Fermi surfaces (hole and electron pockets) and has presented fascinating physics related to superconductivity[1]. FeSe has unique phase diagrams: the superconductivity coexists with nematicity at ambient pressure without magnetism, while the superconductivity coexists with magnetism under pressure instead of nematicity[2]. Such unique features are maintained for S substitution to some extent[3]. At ambient pressure, the nematicity in FeSe$_{1-x}$S$_x$ extends up to the nematic quantum critical point (QCP) ($x_c \simeq 0.17$)[4,5]. Superconducting (SC) transition temperature, $T_c$ of pure FeSe (=9 K) is maintained for S substitution up to $x_c$ and decreases by half when crossing the QCP[6].

Among a variety of topics associated with superconductivity, the topologically protected nodal planes, referred to as Bogoliubov Fermi surfaces (BFSs)[7,8] have been paid much attention in both theoretical and experimental aspects for a heavily S-substituted regime over $x>x_c$. BFSs are attainable in a multiband system under the condition of a strong spin-orbital coupling and time-reversal symmetry breaking (TRSB). An anomalous residual density of states (DOS) in the tetragonal phase with $C_4$ symmetry has raised attention to the study of BFSs. Survival of large DOS in the SC state was found from specific heat[6,9] and scanning tunneling spectroscopy (STS)[5,10] measurements. Following these experimental results, C. Setty et al. have presented a theoretical model based on BFSs in the SC state[8,11], and calculations derived from this model have reproduced these experimental results. The TRSB required for BFSs has been suggested by muon spin relaxation measurements[12]. In recent angular-resolved photoemission spectroscopy (ARPES) experiments[13], heavily S-doped FeSe has exhibited wide nodal regions with $C_2$ symmetry in the SC state, despite the tetragonal phase with $C_4$ symmetry. Y. Cao et al. proposed a microscopic model that includes intraband spin-singlet pairing, interband nonunitary spin-triplet pairing, and ferromagnetic exchange coupling[14]. They found that such BFSs with $C_2$ symmetry can be reproducible as well as other pairing symmetry by tuning gap parameters. Recent NMR studies on FeSe$_{1-x}$S$_x$ ($x=0.18$)[15] have shown an unusual enhancement of low-energy spin fluctuations, namely an unusual upturn of the relaxation rate divided by temperature ($1/T_1T$) as temperature decreases to nearly zero, which upholds the strong Bogoliubov quasiparticle interactions in addition to the presence of BFSs. In the follow-up theoretical paper, Y. Cao et al. calculated the spin susceptibilities for the ultranodal states in a minimal two-band model, where the interband nonunitary spin-triplet pairing is responsible for the BFSs[16]. They concluded that certain scattering between coherent spots/segments on the BFSs can get strongly enhanced, resulting in a robust upturn in the relaxation rate when the interaction is strong. Recently, impurity effect on BFSs has been studied using electron irradiation[17].

To investigate the correlation effect deep in the SC state is of great importance not only for establishing the presence of BFSs but also for offering insights into the pairing symmetry of the system. In the present work, we performed $^{77}$Se-NMR to 100 mK under pressure up to 2.0 GPa using a NiCrAl pressure cell[18].

# EXPERIMENTAL RESULTS

For the study on S-substituted FeSe, the SC state has been investigated at ambient pressure so far[15], while the paramagnetic state was investigated up to several pressures by two NMR groups[19-23]. The SC state of pure FeSe was studied by several NMR works[24-27]. Many of them investigated how the superconductivity breaks under a very high field. In general, the relaxation rate divided by temperature $1/T_1T$ provides a measure of low-energy spin fluctuations.

$$\frac{1}{T_1T} \sim \frac{1}{\omega}\sum_{\mathbf{q}} \mathrm{Im}\chi(\mathbf{q}) \tag{1}$$

where $\chi(\mathbf{q})$ is the wave-number ($\mathbf{q}$)-dependent susceptibility. For $x=0.18$, we found an upturn of $1/T_1T$ with decreasing temperature deep in the SC state at ambient pressure in the same manner as the upturn of $1/T_1T$ toward $T_c$ in the normal state[15]. The upturn of $1/T_1T$ deep in the SC state is extremely rare in the SC compounds. It can hardly be explained by extrinsic effects such as the impurity effect[28], Volovik effect[29], or spatial inhomogeneity[5].

In the case of the Volovik effect, $1/T_1T$ = constant should appear at low $T$ below $T_c$ as well as the impurity effect, and the constant value should depend on the magnetic field[29]. Experimentally, $1/T_1T$ =constant is not observed for $x$=0.18 but for $x$=0.05 and 0.10. However, the value differs by one order of magnitude between them, despite that $T_c$s are almost the same and the applied magnetic field was also the same. Such behavior rules out the possibility of the Volovik effect as well as the impurity effect. The upturn of $1/T_1T$ below $T_c$ is attributed to the scattering between segments on BFSs[16] and gives evidence of the presence of BFSs as discussed in the following contents. To study how external parameters such as pressure influence quasiparticle interactions is of great importance to establish the presence of BFSs. We performed $^{77}$Se-NMR experiments under pressure and found that application of pressure up to 2.0 GPa weakens the scattering on BFSs.

Figure 1 shows an overview of $1/T_1T$ under pressure, which is the main focus of this study. In the normal state, an upturn toward $T_c$ is evident at ambient pressure, but it becomes ambiguous at 1.0 GPa, and $1/T_1T$ becomes almost constant at 2.0 GPa. Interestingly, similar suppression of the upturn under pressure is also observed in the SC state below $T_c$. We will discuss the suppression under pressure in the following section.

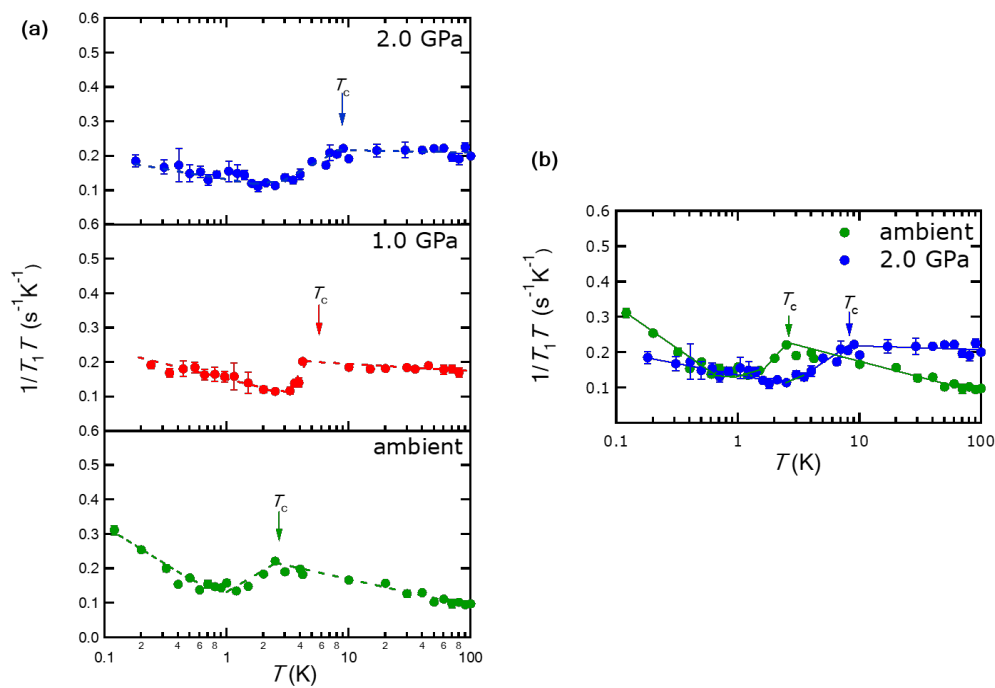

Fig. 1 (a) Temperature ($T$) dependence of relaxation rate ($1/T_1$) divided by $T$, $1/T_1T$, obtained from $^{77}$Se-NMR experiments at several pressure levels on a single crystal of FeSe$_{1-x}$S$_x$ ($x$=0.18). The magnetic field was applied parallel to the FeSe plane. Superconducting transition temperature $T_c$ was measured at 6.02 T by AC susceptibility measurements using a network analyzer. Dashed lines are guides to the eye. (b) Comparison of $1/T_1T$ at 2.0 GPa with that at ambient pressure. Solid lines are guides to the eye.

Figure 2(a) shows the NMR spectra at ambient pressure, 1.0 GPa, and 2.0 GPa, resulting from the FFT of the spin-echo signal. For all three pressure levels, double-peak structure corresponding to the nematic order[30] was not observed throughout the measured temperature range, which is consistent with the results obtained by other techniques. The double-peak structure has been detected at the substitution level below $x_c$ at ambient pressure. Furthermore, disappearance of the echo signal corresponding to AFM ordering was not observed for $x$=0.18 up to 2.0 GPa in contrast to the case of $x$=0.12 at 3.9 GP where AFM order appears[22]. These results indicate that FeSe$_{1-x}$S$_x$ ($x$=0.18) is located in the paramagnetic C$_4$ state up to 2 GPa. Figure 2(b) shows $K$ obtained from NMR spectra in Fig. 2(a). In all three pressure levels, $K$ decreases gently with decreasing temperature and becomes constant at lower temperature regions. Additionally, $K$ also decreases slightly with increasing pressure. The tendency is consistent with that for $x$=0.12[22], 0.15, and 0.29[23].

The Knight shift in NMR experiments can be expressed as the sum of spin part and orbital part, i.e.

$$K = K_{\text{spin}} + K_{\text{orb}}, \qquad (2)$$

in which $K_{\text{spin}}$ is proportional to the density of states (DOS) and the uniform spin susceptibility $\chi(0)$. At low temperatures, both $K_{\text{spin}}$ and $K_{\text{orb}}$ are $T$-independent, and cannot be obtained independently from experiments. The two components are usually separated from each other by the theoretical estimation of $K_{\text{orb}}$. The decrease of $K$ due to pressure application should be attributed to that of $K_{\text{orb}}$, since the DOS of two-dimensional electron systems is constant and proportional to $m/\hbar^2$, where $m$ is the effective mass and $\hbar$ is the Planck constant divided by $2\pi$, and consequently, $K_{\text{spin}}$ should be constant.

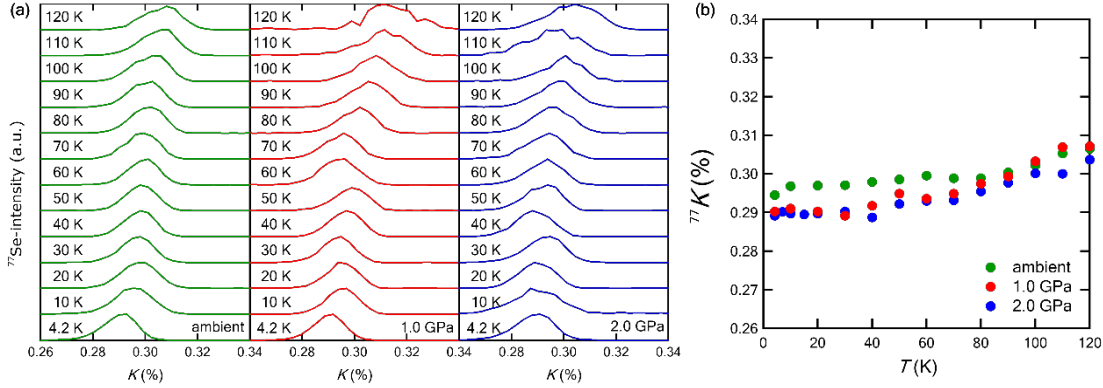

Fig. 2 (a) NMR spectra for FeSe$_{1-x}$S$_x$ ($x$=0.18) at ambient pressure, 1.0 GPa, and 2.0 GPa, resulting from the FFT of the spin-echo signal. (b) Knight shift ($K$) obtained from the peak positions of NMR spectra shown in 2(a).

## DISCUSSION

### Upturn of $1/T_1T$ at ambient pressure ($T>T_c$)

The upturn of $1/T_1T$ toward $T_c$ in the normal state originates from spin fluctuations with $\mathbf{q} \neq 0$, as seen from the comparison of $T$ dependence between $K_{\text{spin}}$ and $1/T_1T$. Similarly to iron-based pnictides, spin fluctuations are associated with the topological configuration of electron and hole pockets and interband coupling. The enhancement of $\chi(\mathbf{q})$ is expected at $\mathbf{q} \sim (\pi, 0)$ in the nematic phase as well as in the tetragonal phase as shown in Fig. 3, owing to the interband coupling between electron and hole pockets, namely the $\mathbf{q} = (\pi, 0)$ nesting, as shown in Fig. 4(a). The enhancement of $\chi(\mathbf{q})$ at $\mathbf{q} \sim (\pi, 0)$ was observed for pure FeSe and 7%-S substituted FeSe in neutron inelastic scattering measurements[31-34].

### Pressure-induced Lifshitz transition

The upturn of $1/T_1T$ above $T_c$ is suppressed with increasing pressure, and $1/T_1T$ becomes a constant at 2.0 GPa. The behavior is not specific to $x$=0.18 but is common to all substitution levels. At a high-pressure region, a pressure-induced Lifshitz transition occurs as suggested from both theoretical and experimental aspects[20, 22, 35-38]. Figure 3 gives an overview of three different substitution levels $x$=0.05, 0.12[22], and 0.18, crossing the nematic QCP. The data for $x$=0.18 is the same as those shown in Figs. 1(a) and 1(b), although the upturn at ambient pressure looks small on this scale. Similar behavior has also been observed by another NMR study for $x$=0.20, and Fermi-liquid behavior was suggested in their study[23].

According to the theoretical investigation[35], reconstruction of the Fermi surfaces or predominant $\mathbf{q}$, namely Lifshitz transition, occurs at around 2 GPa due to the emergence of $d_{xy}$ hole pocket at $\mathbf{k} \sim (\pi, \pi)$ as shown in Fig. 4(b), which could cause the $\mathbf{q}=(0, \pi)$ nesting between electron and hole pockets at a high pressure regime, as expected in the $P$-$T$ phase diagram of Fe(Se,S)[3]. At the crossover region where the dominant $\mathbf{q}$ reconstructs and the nesting changes, spin fluctuations become tentatively $T$-independent before forming a new nesting condition.

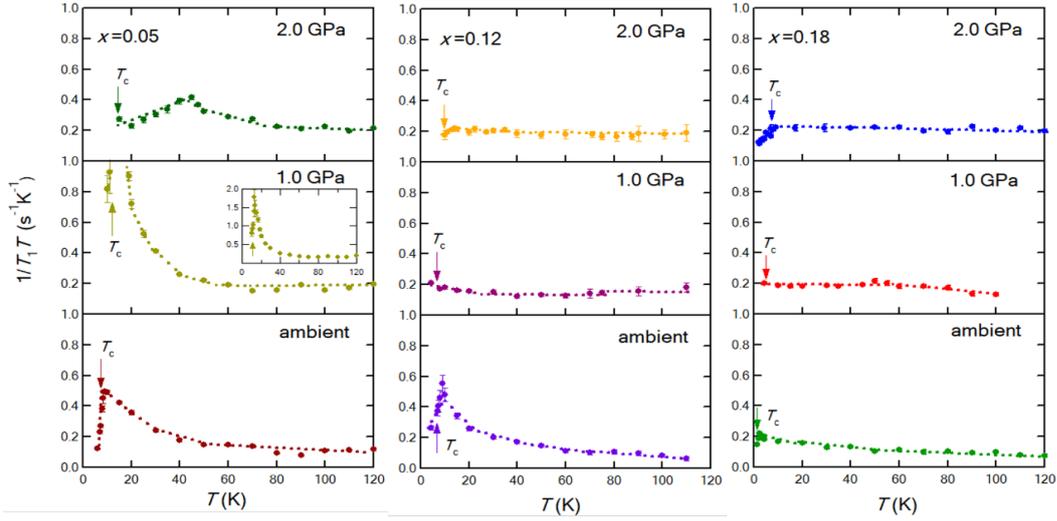

Fig. 3 An overview of $1/T_1T$ for three different substitution levels, $x$=0.05, 0.12[20], 0.18, crossing the nematic QCP. The data of the inset is the same with those of the main panel but displayed in different scale.

Notably, an extremely strong peak is observed for $x$=0.05 at 1.0 GPa. However, it is strongly suppressed at 2.0 GPa, and another peak corresponding to AFM ordering appears at around 40 K. This phenomenon is explained by the Lifshitz transition occurring at a lower pressure region below 2 GPa. The linewidth broadens a little at around 40 K. For $x$=0.12, the AFM ordering is detected at a higher pressure as the loss of signal[22], as shown in Supplementary Fig. 2.

## Spin correlation at 2.0 GPa ($T>T_c$)

The suppression of the upturn under pressure implies that the interaction between electrons becomes weak under pressure. The interaction at 2.0 GPa can be estimated given that $1/T_1T$ is almost $T$-independent and obeys the Korringa relation:

$$\frac{1}{T_1T} = \frac{4\pi k_B}{\hbar}\left(\frac{\gamma_n}{\gamma_e}\right)^2 K_{\rm spin}^2 K(\alpha). \tag{3}$$

The interaction $U$ is included in $\alpha$ as $\alpha=U\chi_0$ where $\chi_0$ is the uniform susceptibility of free electrons. $K(\alpha)$ provides a measure of electron correlation[15]:

$$K(\alpha) = \frac{(1-\alpha)^2}{\left\langle\left(1-\frac{\alpha\chi_0(\mathbf{q})}{\chi_0}\right)^2\right\rangle_{\rm FS}}, \tag{4}$$

where $\left\langle\left(1-\frac{\alpha\chi_0(q)}{\chi_0}\right)^2\right\rangle_{\rm FS}$ represents the average over $\mathbf{q}$ connecting two points on the Fermi surfaces, and $\chi_0(\mathbf{q})$ is the $\mathbf{q}$-dependent susceptibility without the interaction. For ferromagnetic metals, $K(\alpha)<1$, whereas for antiferromagnetic metals, $K(\alpha)>1$. For free electrons, $K(\alpha)=1$ and $\alpha=0$. For pure FeSe, $K_{\rm orb}\approx0.23\%$ and $K_{\rm spin}=0.06\%$ were estimated from the $K$-$\chi$ plot[39], while $K_{\rm orb}/K_{\rm spin}\approx5$ estimated from RPA calculations[40] leads to $K_{\rm orb}\approx0.25\%$ and $K_{\rm spin}=0.05\%$. For $x$=0.12, $K_{\rm orb}\approx0.26\%$ and $K_{\rm spin}=0.03\%$ were estimated from the $K$-$\chi$ plot[22]. These results imply that $K$ is almost insensitive to S-substitution levels and is estimated to be approximately 0.3% for $x$ up to ~0.29[23]. As for pressure dependence, $K$ decreases only 0.01% by applying pressure of 2.0 GPa, as shown in Fig. 2(b). The decrease is approximately the same for $x$=0.12, implying that $K$ is also almost insensitive to pressure. We estimate $K(\alpha)\approx15$ from $1/T_1T\approx0.21$ assuming $K_{\rm spin}=0.03\%$ at 2.0 GPa for $x$=0.18. The results suggest that AFM fluctuations still remain even at 2.0 GPa, although AFM fluctuations or the upturn toward $T_c$ is strongly suppressed with increasing pressure.

## Theoretical model based on BFSs with $C_2$ symmetry ($T<T_c$)

Similarly to the normal state, we suggested that $1/T_1T$ deep in the SC state would be explained by a nesting scenario because the upturn below $T_c$ resembles the upturn above $T_c$.[15] In this nesting scenario, BFSs are supposed to exist for both hole and electron pockets. However, it is unclear whether BFSs exist on the electron pocket at the present stage, because the gap closing has been observed only for a hole pocket at Γ point from the Laser ARPES measurements[13].

Contrary to the scenario above, the recent theoretical investigation[16] pointed out that the upturn below $T_c$ is explainable only by BFSs appearing at hole pockets as shown in Fig. 4(c). This theoretical model assumes two hole pockets and BFSs with $C_2$ symmetry at Γ point. The wave number $\mathbf{q}\sim(0.4\pi, 0)$ corresponding to a span between two BFSs segments contributes to an enhancement of $\chi(\mathbf{q})$ under the strong Hubbard-type interaction $U$. The BFSs segments colored in purple have the character of significant interband spin-triplet particle-hole mixing and moderate intraband spin-singlet particle-hole mixing. $1/T_1T$ was calculated under the random phase approximation (RPA) for two cases where $\boldsymbol{B}$ is applied parallel (//) and perpendicular (⊥) to the quantization axis ($z$-axis) of the spin-triplet[16]. In both cases, the upturn of $1/T_1T$ is reproduced. The case for $\boldsymbol{B}\perp z$ causes a significant upturn at low temperatures and seems to reproduce the experimental results better than that for $\boldsymbol{B}//z$[16].

The upturn of $1/T_1T$ originates from two factors: the nesting between BFSs segments and the interaction $U$. However, it is difficult to specify which contribution is larger from the experiments alone. According to the theoretical calculation of $1/T_1T$ under RPA mentioned above[16], the upturn is hardly expected if $U$ is zero. This fact suggests that the upturn is sensitive to $U$. As shown in Figs. 1(a) and 1(b), the upturn of $1/T_1T$ below $T_c$ is suppressed at 2.0 GPa compared to that at ambient pressure, implying that $U$ becomes weak but nonzero under pressure. The result seems reasonable considering $K(\alpha) \simeq 15$ at 2.0 GPa in the normal state because $K(\alpha) \simeq 1$ is expected for $U=0$. The theoretical model would also explain why the $T$ dependence of $1/T_1T$ is different across $T_c$: this is clearly seen from the data at 2.0 GPa where $1/T_1T$ =constant above $T_c$ whereas $1/T_1T$ exhibits an upturn below $T_c$. This change would be because the nesting properties change across $T_c$. The predominant $\mathbf{q}$ would change from $\mathbf{q} = (\pi, 0)$ above $T_c$ to $\mathbf{q}\sim(0.4\pi, 0)$ below $T_c$, and $\chi_0(\mathbf{q})$ in the SC state would be different from that determined by the nesting between normal electrons.

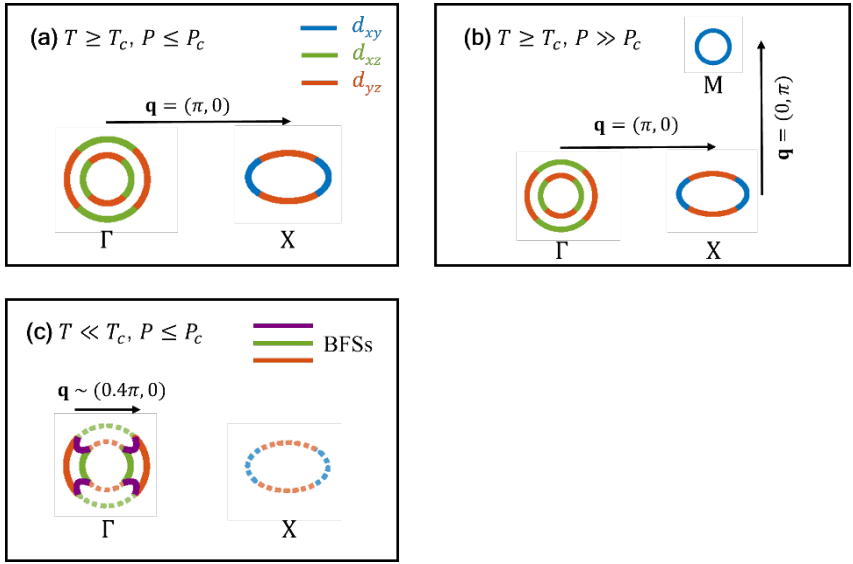

Fig. 4 (a) Original Fermi surfaces (OFSs) in the normal state at ambient pressure. (b) OFSs in the normal state under pressure with another hole pocket emerging at point M, $\boldsymbol{k} = (\pi, \pi)$. (c) Bogoliubov Fermi surfaces (BFSs) in the SC state[16]. OFSs in the normal metallic state are shown by dashed curves in (c) for comparison. BFSs in the superconducting state are plotted by solid curves. The segments colored in red and green represent BFSs with the character of small intraband spin-singlet particle-hole mixing, and the segments colored in purple represent BFSs with the character of significant interband spin-triplet particle-hole mixing.

Recently, another theoretical model based on the pairing mediated by soft charge-nematic fluctuations has been proposed[41, 42]. In this model, the strength depends on the location of a fermion along the Fermi surface, which induces nodal areas below $T<T_c$. Our experimental results exhibit the upturn of $1/T_1T$ down to T≃$0.02T_c$. It is a future problem whether this model can reproduce the upturn of $1/T_1T$ and its suppression under pressure.

## Conclusion

We have demonstrated that the enhancement of $\chi(\mathbf{q})$, namely the upturn of $1/T_1T$, at ambient pressure observed deep in the SC state is suppressed by the application of pressure, which implies that the scattering between the segments on nodal areas becomes weak but nonzero under pressure. The correlation on the nodal areas below $T_c$ seems stronger than that between electron and hole pockets above $T_c$, considering the results at 2.0 GPa. These experimental results are explainable by the recent theoretical model of BFSs with $C_2$ symmetry at Γ point. The upturn of $1/T_1T$ at ambient pressure is attributed to the enhancement of $\chi(\mathbf{q})$ under strong interaction at $\mathbf{q}$≃$(0.4\pi, 0)$, a span between two BFSs segments with the character of significant interband spin-triplet particle-hole mixing. These experimental facts contribute to establish the presence of the Bogoliubov quasiparticle interactions along with the presence of BFSs themselves. This study advances the understanding of interactions on the BFSs and offers a crucial step towards unraveling the nature of the pairing symmetry of the system.

## Methods

In this study, [77]Se-NMR measurements of 18% S-substituted FeSe were conducted under hydrostatic pressure up to 2.0 GPa down to 100 mK. We used the sole single crystal with a size of 1.0×1.0×0.5 mm³ that was used in the previous measurements[15]. Nuclear magnetic relaxation time ($T_1$) was measured by using the saturation-recovery method and evaluated by the exponential fitting of the recovery curve. Knight shift ($K$) was determined from the peak frequency of the fast Fourier transform (FFT) of the NMR signal. A magnetic field of 6.02 T was applied parallel to the FeSe plane ($\boldsymbol{B}$//ab) to avoid vortices. We used a dilution refrigerator made by Bluefors for measurements at low temperatures below 1 K. Temperature fluctuations were suppressed within an accuracy of 0.01K by PID control. The heating due to the NMR pulses was not observed within the experimental accuracy. These experimental conditions are the same with Ref 15. To perform NMR measurements under pressure, we used a NiCrAl piston-cylinder pressure cell[18] and monitored the pressure inside by ruby fluorescence measurements. For this purpose, we inserted an optical fiber with ruby powders on one end into the pressure cell together with an NMR coil. $T_c$ was determined from AC susceptibility measurements utilizing the NMR tank circuit. The results measured at zero field and 6.02 T were shown in Supplementary Fig. 1.

## Acknowledgements

The authors would like to thank S. Nagasaki and T. Takahashi for their experimental support. The present work was supported by JST SPRING, Grant Number JPMJSP2110. This work was partly supported by Grants-in-Aid for Scientific Research (KAKENHI Grant Nos. JP22H00105 and JP19H00648) and by Innovative Areas "Quantum Liquid Crystals" (No. JP19H05824) from the Japan Society for the Promotion of Science.

## Author contributions

N.F. designed the NMR experiments. Z.Y., K.N., and K.I. carried out the NMR measurements. K.M., Y. Mizukami, S.K., Y. Matsuda, and T.S. synthesized the samples and performed the chemical analysis. X.S. operated the dilution refrigerator to cool down to 100 mK under the instruction of Y. Uwatoko.

## Data avaiavirity

The data that support the findings of this study are available from the corresponding author upon reasonable request.

# References

1. T. Shibauchi, T. Hanaguri, and Y. Matsuda, Exotic superconducting states in FeSe-based materials, J. Phys. Soc. Jpn. **89**, 102002 (2020).
2. J. P. Sun, K. Matsuura, G. Z. Ye, Y. Mizukami, M. Shimozawa, K. Matsubayashi, M. Yamashita, T. Watashige, S. Kasahara, Y. Matsuda, J. -Q. Yan, B. C. Sales, Y. Uwatoko, J. -G. Cheng & T. Shibauchi, Dome-shaped magnetic order competing with high-temperature superconductivity at high pressures in FeSe, Nat. Commun. **7**, 12146 (2016).
3. K. Matsuura *et al.*, Maximizing by tuning nematicity and magnetism in $FeSe_{1-x}S_x$ superconductors, Nat. Commun. **8**, 1143 (2017).
4. S. Hosoi, K. Matsuura, K. Ishida, H. Wang, Y. Mizukami, T. Watashige, S. Kasahara, Y. Matsuda, and T.Shibauchi, Nematic quantum critical point without magnetism in $FeSe_{1-x}S_x$ superconductors, Proc. Natl Acad. Sci. U.S.A. **113**, 8139 (2016).
5. T. Hanaguri, K. Iwaya, Y. Kohsaka, T. Machida, T. Watashige, S. Kasahara, T. Shibauchi, and Y. Matsuda, Two distinct superconducting pairing states divided by the nematic end point in $FeSe_{1-x}S_x$, Sci. Adv. **4**, eaar6419 (2018)..
6. Y. Mizukami, M. Haze, O. Tanaka, K. Matsuura, D. Sano, J. Böker, I. Eremin, S. Kasahara, Y. Matsuda, and T. Shibauchi, Unusual crossover from Bardeen-Cooper-Schrieffer to Bose-Einstein-condensate superconductivity in iron chalcogenides, Commun. Phys. **6**, 183 (2023).
7. D. F. Agterberg, P. M. R. Brydon, and C. Timm, Bogoliubov Fermi surfaces in superconductors with broken time-reversal symmetry, Phys. Rev. Lett. **118**, 127001 (2017).
8. C. Setty, S. Bhattacharyya, Y. Cao, A. Kreisel, and P. J. Hirschfeld, Topological ultranodal pair states in iron-based superconductors, Nat. Commun. **11**, 523 (2020).
9. Y. Sato, S. Kasahara, T. Taniguchi, X. Xing, Y. Kasahara, Y. Tokiwa, Y. Yamakawa, H. Kontani, T. Shibauchi, and Y. Matsuda, Abrupt change of the superconducting gap structure at the nematic critical point in $FeSe_{1-x}S_x$, Proc. Natl. Acad. Sci. U.S.A. **115**, 1227 (2018).
10. P. K. Nag, K. Scott, V. S. de Carvalho, J. K. Byland, X. Yang, M. Walker, A. G. Greenberg, P. Klavins, E. Miranda, A. Gozar, V. Taufour, R. M. Fernandes, E. H. da Silva Neto, Superconductivity mediated by nematic fluctuations in tetragonal $FeSe_{1-x}S_x$, 2024, https://doi.org/10.48550/arXiv.2403.00615
11. C. Setty, Y. Cao, A. Kreisel, S. Bhattacharyya, and P. J. Hirschfeld, Bogoliubov Fermi surfaces in spin-1/2 systems: Model Hamiltonians and experimental consequences, Phys. Rev. B **102**, 064504 (2020).
12. K. Matsuura *et al.*, Two superconducting states with broken time-reversal symmetry in $FeSe_{1-x}S_x$, Proc. Natl. Acad. Sci. U.S.A. **120**, e2208276120 (2023).
13. T. Nagashima, T. Hashimoto, S. Najafzadeh, S. ichiro Ouchi, T. Suzuki, A. Fukushima, S. Kasahara, K. Matsuura, M. Qiu, Y. Mizukami, K. Hashimoto, Y. Matsuda, T. Shibauchi, and K. Okazaki, Discovery of nematic Bogoliubov Fermi surface in an iron-chalcogenide superconductor, Research Square, 2022, https://doi.org/10.21203/rs.3.rs-2224728/v1.
14. Y. Cao, C. Setty, L. Fanfarillo, A. Kreisel, and P. J. Hirschfeld, Microscopic origin of ultranodal superconducting states in spin-1/2 systems, Phys. Rev. B **108**, 224506 (2023).
15. Z. Yu, K. Nakamura, K. Inomata, X. Shen, T. Mikuri, K. Matsuura, Y. Mizukami, S.Kasahara, Y.Matsuda, T. Shibauchi, Y. Uwatoko, and N. Fujiwara, Spin fluctuations from Bogoliubov Fermi surfaces in the superconducting state of S-substituted FeSe, Commun. Phys. **6**, 175 (2023).
16. Y. Cao, C. Setty, A. Kreisel, L. Fanfarillo, and P. J. Hirschfeld, Spin fluctuations in the ultranodal superconducting state of Fe(Se,S). Phys. Rev. B **110**, L020503 (2024).
17. T. Nagashima, K. Ishihara, K. Imamura, M. Kobayashi, M. Roppongi, K. Matsuura, Y. Mizukami, R. Grasset, M. Konczykowski, K. Hashimoto, and T. Shibauchi, Lifting of gap nodes by disorder in tetragonal $FeSe_{1-x}S_x$ superconductors, Phys. Rev. Lett. **133**, 156506 (2024).
18. N. Fujiwara, T. Matsumoto, K. Koyama-Nakazawa, A. Hisada, and Y. Uwatoko, Fabrication and efficiency evaluation of a hybrid NiCrAl pressure cell up to 4 GPa, Rev. Sci. Instrum. **78**, 073905 (2007).

19. P. Wiecki, K. Rana, A. E. Böhmer, Y. Lee, S. L. Bud'ko, P. C. Canfield, and Y. Furukawa, Persistent correlation between super-conductivity and antiferromagnetic fluctu-ations near a nematic quantum critical point in FeSe$_{1-x}$S$_x$, Phys. Rev. B **98**, 020507(R) (2018).
20. T. Kuwayama, K. Matsuura, Y. Mizukami, S. Kasahara, Y. Matsuda, T. Shibauchi, Y. Uwatoko, and N. Fujiwara, $^{77}$Se-NMR study under pressure on 12%-S doped FeSe, J. Phys. Soc. Jpn. **88**, 033703 (2019).
21. K. Rana, L. Xiang, P. Wiecki, R. A. Ribeiro, G. G. Lesseux, A. E. Böhmer, S. L. Bud'ko, P. C. Canfield, and Y. Furukawa, Impact of nematicity on the relationship between antiferromagnetic fluctuations and super-conductivity in FeSe$_{0.91}$S$_{0.09}$ under pressure, Phys. Rev. B **101**, 180503 (2020).
22. T. Kuwayama, K. Matsuura, J. Gouchi, Y. Yamakawa, Y. Mizukami, S. Kasahara, Y. Matsuda, T. Shibauchi, H. Kontani, Y. Uwatoko, and N. Fujiwara, Pressure-induced reconstitution of Fermi surfaces and spin fluctuations in S-substituted FeSe, Sci. Rep. **11**, 17265 (2021).
23. K. Rana, D. V. Ambika, S. L. Bud'ko, A. E. Böhmer, P. C. Canfield, and Y. Furukawa, Interrelationships between nematicity, antiferromagnetic spin fluctuations, and superconductivity: Role of hotspots in FeSe$_{1-x}$S$_x$ revealed by high pressure $^{77}$Se NMR study, Phys. Rev. B **107**, 134507 (2023).
24. A. Shi, T. Arai, S. Kitagawa, T. Yamanaka, K. Ishida, A. E. Böhmer, C. Meingast, T. Wolf, M. Hirata, and T. Sasaki, Pseudogap behavior of the nuclear spin–lattice relaxation rate in FeSe probed by $^{77}$Se-NMR, J. Phys. Soc. Jpn. **87**, 013704 (2018).
25. S. Molatta, D. Opherden, J. Wosnitza, L. Opherden, Z. T. Zhang, T. Wolf, H. v. Löhneysen, R. Sarkar, P. K. Biswas, H.-J. Grafe, and H. Kühne, Superconductivity of highly spin-polarized electrons in FeSe probed by $^{77}$Se NMR, Phys. Rev. B **104**, 014504 (2021).
26. I. Vinograd, S. P. Edwards, Z. Wang, T. Kissikov, J. K. Byland, J. R. Badger, V. Taufour, and N. J. Curro, Inhomogeneous Knight shift in vortex cores of superconducting FeSe, Phys. Rev. B **104**, 014502 (2021).
27. J. Li, B. L. Kang, D. Zhao, B. Lei, Y. B. Zhou, D. W. Song, S. J. Li, L. X. Zheng, L. P. Nie, T. Wu, and X. H. Chen, $^{77}$Se -NMR evidence for spin-singlet superconductivity with exotic superconducting fluctuations in FeSe, Phys. Rev. B **105**, 054514 (2022).
28. Y. Bang, H. Y. Choi, and H. Won, Impurity effects on the ±s-wave state of the iron-based superconductors, Phys. Rev. B **79**, 054529 (2009).
29. Y. Bang, Volovik effect on NMR measurements of unconventional super-conductors, Phys. Rev. B **85**, 104524 (2012).
30. S-H. Baek, D. V. Efremov, J. M. Ok, J. S. Kim, J. van den Brink & B. Büchner, Orbital-driven nematicity in FeSe, Nat. Mater. **14**, 210 (2015).
31. M. C. Rahn, R. A. Ewings, S. J. Sedlmaier, S. J. Clarke, and A. T. Boothroyd, Strong (π, 0) spin fluctuations in β−FeSe observed by neutron spectroscopy, Phys. Rev. B **91**, 180501 (2015).
32. Q. Wang, Y. Shen, B. Pan, Y. Hao, M. Ma, F. Zhou, P. Steffens, K. Schmalzl, T. R. Forrest, M. Abdel-Hafiez, X. Chen, D. A. Chareev, A. N. Vasiliev, P. Bourges, Y. Sidis, H. Cao, and J. Zhao, Strong interplay between stripe spin fluctuations, nematicity and superconductivity in FeSe, Nature Mater. **15**, 159–163 (2016).
33. Q. Wang, Y. Shen, B. Pan, X. Zhang, K. Ikeuchi, K. Iida, A. D. Christianson, H. C. Walker, D. T. Adroja, M. Abdel-Hafiez, X. Chen, D. A. Chareev, A. N. Vasiliev, and J. Zhao, Magnetic ground state of FeSe, Nat. Comm. **7**, 12182 (2016).
34. M. Ma, P. Bourges, Y. Sidis, J. Sun, G. Wang, K. Iida, K. Kamazawa, J. T. Park, F. Bourdarot, Z. Ren, and Y. Li, Ferromagnetic inter-layer coupling in FeSe$_{1-x}$S$_x$ superconductors revealed by inelastic neutron scattering, 2024, https://doi.org/10.48550/arXiv.2407.05548.
35. Y. Yamakawa, and H. Kontani, Nematicity, magnetism, and superconductivity in FeSe under pressure: Unified explanation based on the self-consistent vertex correction theory. Phys. Rev. B **96**, 144509 (2017).
36. P. Reiss, D. Graf, A. A. Haghighirad, W. Knafo, L. Drigo, M. Bristow, A. J. Schofield & A. I. Coldea, Quenched nematic criticality and two superconducting domes in an iron-based superconductor, Nat. Phys. **16**, 89-94 (2020).
37. Z. Zajicek, P. Reiss, D. Graf, J. C. A. Prentice, Y. Sadki, A. A. Haghighirad, and A. I. Coldea, Unveiling the quasiparticle behaviour in the pressure-induced high-$T_c$ phase of an iron-chalcogenide superconductor, Npj Quantum Materials **9**, 52 (2024).

38. P. Reiss, A. McCollam, Z. Zajicek, A. A. Haghighirad & A. I. Coldea, Collapse of metallicity and high-$T_c$ superconductivity in the high-pressure phase of FeSe$_{0.89}$S$_{0.11}$, Npj Quantum Materials **9**, 73 (2024).
39. See Supplemental Material at https://journals.aps.org/prx/supplemental/10.1103/PhysRevX.10.011034/ for more information of the $K$-$\chi$ plot and Knight-shift.
40. R. Zhou, D. D. Scherer, H. Mayaffre, P. Toulemonde, M. Ma, Y. Li, B. M. Andersen, and M.-H. Julien, Singular magnetic anisotropy in the nematic phase of FeSe, Npj Quantum Materials, **5**, 93 (2020).
41. K. R. Islam and A. Chubukov, Unconventional superconductivity near a nematic instability in a multiorbital system, npj Quantum Materials **9**, 28 (2024).
42. K. R. Islam and A. Chubukov, Unconventional superconductivity mediated by nematic fluctuations in a multiorbital system: Application to doped FeSe, Phys. Rev. B **111** 094503(2025).

# Supplementary Material-Quasiparticle interaction originating from Bogoliubov Fermi Surfaces under pressure in 18%-S substituted FeSe studied via NMR

Zhongyu Yu[1], Xiaoling Shen[2,*], Koya Nakamura[1,†], Kazuya Inomata[1,§], Kohei Matsuura[3,‡], Yuta Mizukami[3,**], Shigeru Kasahara[4,††], Yuji Matsuda[4], Takasada Shibauchi[3], Yoshiya Uwatoko[2], and Naoki Fujiwara[1,§§]

[1]Graduate School of Human and Environmental Studies, Kyoto University, Yoshida-Nihonmatsu-cho, Sakyo-ku, Kyoto, 606-8501, Japan
[2]Institute for Solid State Physics, University of Tokyo, 5-1-5 Kashiwanoha, Kashiwa, 277-8581, Chiba, Japan
[3]Department of Advanced Materials Science, University of Tokyo, 5-1-5 Kashiwanoha, Kashiwa, 277-8561, Chiba, Japan
[4]Department of Physics, Kyoto University, Kitashirakawa Oiwake-cho, Sakyo-ku, Kyoto, 606-8502, Japan
*Present address: Key Laboratory of Artificial Structures and Quantum Control, School of Physics and Astronomy, Shanghai Jiao Tong University, Shanghai 200240, China
†Present address: NEC Platforms, Ltd., 1753-1 Shimonumabe, Nakahara-ku, Kawasaki, Kanagawa, 211-8666, Japan
§Present address: DeNA Co., Ltd., 6-30-15 Hommachi, Shibuya-ku, Tokyo, 151-0071, Japan
‡Present address: Department of Applied Physics, University of Tokyo, 7-3-1 Hongo, Bunkyou-ku, Tokyo, 113-8656, Japan
**Present address: Graduate School of Science, Department of Physics, Tohoku University, Sendai, Miyagi, 980-8578, Japan
††Present address: Research Institute for Interdisciplinary Science, Okayama University, Okayama, 700-8530, Japan
§§Contact author: fujiwara.naoki.7e@kyoto-u.ac.jp

# I. DETERMINATION OF $T_c$ FROM THE AC SUSCEPTIBILITY

The AC susceptibility measurements for $x$=0.18 were conducted to determine $T_c$ by using the tank circuit attached to the head of an NMR probe at zero field and 6.02T. The magnetic field was applied parallel to the FeSe plane (**B**//ab) similarly to the $T_1$ measurements.

The resonance frequency of the tank circuit ($f_r$) was measured using a commercially available network analyzer. The frequency $f_r$ is related to the AC susceptibility χ as $f_r = 1/\sqrt{LC(1+4\pi\chi)}$, where $C$ and $L$ represent the capacitance of a variable capacitor and the inductance of a coil wound onto the sample, respectively. The frequency $f_r$ increases gradually with decreasing temperature due to a gradual decrease in $L$ during the cooling process, and changes drastically at $T_c$ owing to the Meissner effect. We determined $T_c$s from the crossing points of the dashed lines. $T_c$s at zero field were 4.0, 6.7 and 12.4 K at ambient pressure, 1.0 and 2.0 GPa, respectively, while those at 6.02 T were 2.8, 6.1 and 9.5 K at ambient pressure, 1.0 and 2.0 GPa, respectively.

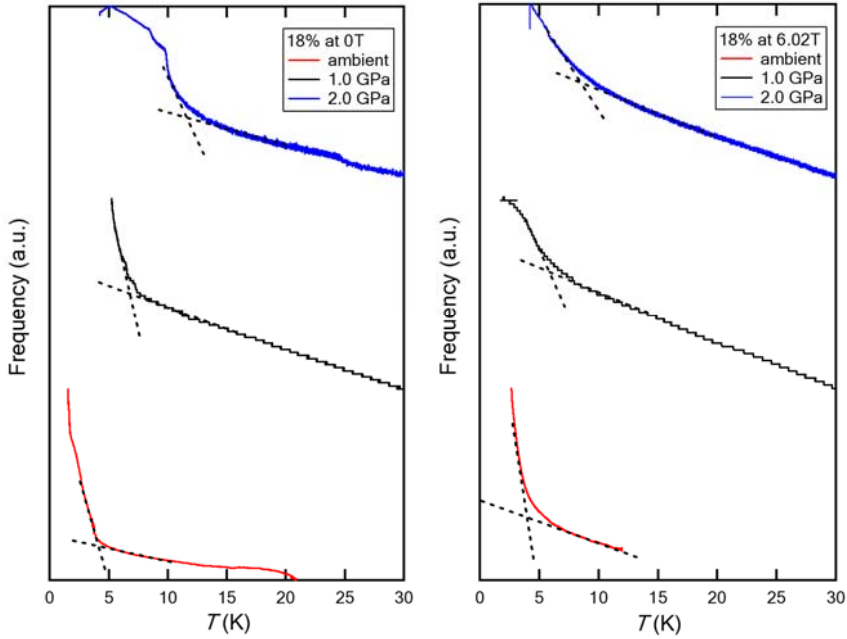

Figure 1: AC susceptibility measurements for FeSe$_{1-x}$S$_x$ ($x$=0.18) using the tank circuit attached on the top of an NMR probe. $T_c$s were determined from the crosspoints of dashes lines. Left and right panels represent the measurements at zero field and 6.02 T, respectively.

# II. EVIDENCE OF ANTIFERROMAGNETIC ORDER UNDER PRESSURE

The appearance of antiferromagnetic (AFM) ordering can be observed from anomaly in linewidth of NMR spectra. Figure 2 shows the linewidth of $x$ = 0.05 and 0.12 [1] at different pressure levels. For $x = 0.12$, NMR signals disappear as AFM order appears at 3.9 GPa, leading to an increase of linewidth, whereas no significant behavior is observed at 3.5 GPa where AFM order is absent. For $x = 0.05$, a hump is observed at 2.0 GPa and develops into a larger peak with further applying pressure to 2.8 GPa. The hump and the peak are considered as a precursor of AFM ordering and the intermediate stage of the development of AFM order, respectively. The NMR signal for $x=0.05$ is lost at $T_c$ accompanied with the increase in linewidth.

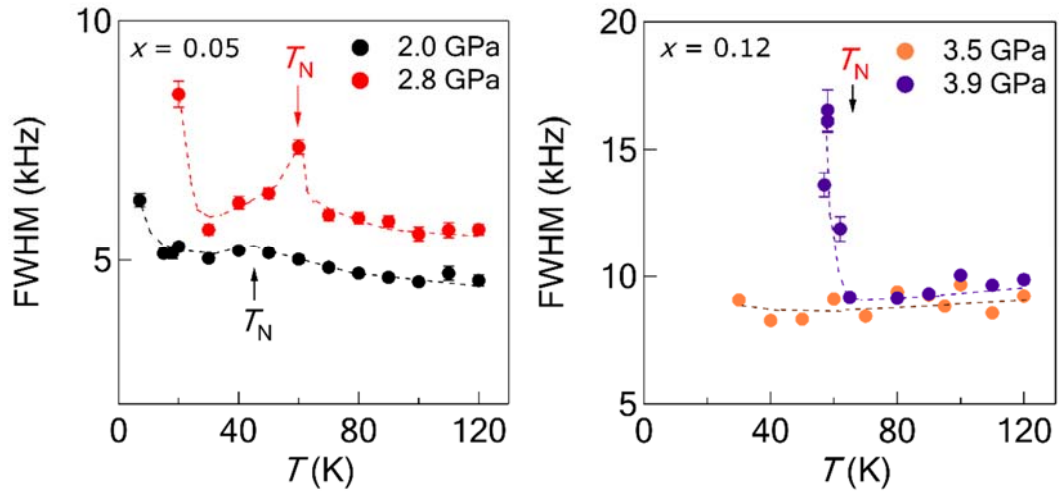

Figure 2: Linewidth of NMR spectra at different pressure levels for $x = 0.05$ and 0.12.

[1] T. Kuwayama, K. Matsuura, Y. Mizukami, S. Kasahara, Y. Matsuda, T. Shibauchi, Y. Uwatoko, and N. Fujiwara, $^{77}$Se-NMR study under pressure on 12%-S doped FeSe, J. Phys. Soc. Jpn. **88**, 033703 (2019).